\documentclass[letterpaper,english,prl,twocolumn]{revtex4}
\usepackage[T1]{fontenc}
\usepackage[latin1]{inputenc}
\usepackage{subfigure}
\usepackage{graphicx}

\makeatletter

\usepackage{babel}
\makeatother
\begin{document}

\preprint{This line only printed with preprint option}

\title{From $E_{2g}$ \textmd{\normalsize }to other modes : Effects of pressure
on electron-phonon interaction in $\mathbf{Mg}\mathbf{B}_{2}$}

\author{Prabhakar P. Singh}

\email{ppsingh@phy.iitb.ac.in}

\affiliation{Department of Physics, Indian Institute of Technology, Powai, Mumbai-
400076, India }

\begin{abstract}
We study the effects of pressure on the electron-phonon interaction
in $MgB_{2}$ using density-functional-based methods. Our results
show that the superconductivity in $MgB_{2}$ vanishes by $100$ GPa,
and then \emph{reappears} at higher pressures. In particular, we find
a superconducting transition temperature {\normalsize $T_{c}\approx2$
K for} $\mu^{*}=0.1$ at a pressure of $137$ GPa. 
\end{abstract}
\maketitle
The superconductivity in $MgB_{2}$ \cite{akimitsu} is intimately
connected with the holes in the $B$ $p_{x(y)}$ and $p_{z}$ bands
\cite{mgb2a,kortus,pps_mgb2} and their coupling to the in-plane $E_{2g}$
phonon mode \cite{bohnen,kong,choi1,yildirim}. Several experiments
on $MgB_{2}$ under pressure \cite{monteverde,tomita,deemyaad} show
a reduction in superconducting transition temperature $T_{c}$ at
a rate of -$1.1$ K GPa$\,^{-1}$, which can be understood in terms
of a continuous weakening of $E_{2g}$ phonon mode coupling. Based
on this picture, it is estimated that the superconductivity in $MgB_{2}$
will vanish by $90$-$100$ GPa \cite{zhang}. 

Recent experiments on elemental $Li$ \cite{shimuzu,stru} and $Y$
\cite{hamlin} have shown a significant increase in their $T_{c}$
with increase in pressure. For example, the $T_{c}$ of $Y$ changes
from $6$ mK \cite{probst} to $20$ K \cite{hamlin} as the pressure
is increased from ambient to $115$ GPa. In light of these experiments,
it would be interesting to see what happens to the superconducting
properties, especially $T_{c},$ of $MgB_{2}$ beyond $90$-$100$
GPa. 

In this Letter, we report on our density-functional-based study of
the evolution of electron-phonon interaction in $MgB_{2}$ as a function
of pressure up to $150$ GPa. Additionally, we have solved the isotropic
Eliashberg gap equation to obtain $T_{c}$ of $MgB_{2}$ as a function
of pressure. 

Based on our calculations, we find that the increase in pressure hardens
the $MgB_{2}$ lattice and dramatically reduces the contribution of
the $E_{2g}$ phonon mode to the electron-phonon coupling. As a result,
the superconductivity in $MgB_{2}$ vanishes by $100$ GPa, only to
\emph{reappear} at higher pressures. In particular, we find a superconducting
transition temperature $T_{c}\approx2$ K for $\mu^{*}=0.1$ at a
pressure of $137$ GPa. We also find that the $B_{1g}$ phonon mode
completely softens around $A$ symmetry point by $143$ GPa. Before
we describe our results, we briefly outline the computational details
of our calculations. 

We have calculated the electronic structure and the electron-phonon
interaction of $MgB_{2}$ in $P6/mmm$ crystal structure as a function
of pressure up to $150$ GPa. The lattice constants $a$ \emph{}and
\emph{$c$} for different pressures were taken from Ref. \cite{zhang},
which agree well in the range of pressures where experiments have
been reported. The electronic structure was calculated using the full-potential
linear muffin-tin orbital (LMTO) method \cite{savrasov1,savrasov2}
as well as the plane-wave pseudopotential method using PWSCF package
\cite{pwscf}. The phonon dispersion, $\omega_{\mathbf{q}\nu}$, the
phonon linewidth, $\gamma_{\mathbf{q}\nu}$, the phonon density of
states, $F(\omega)$, and the Eliashberg function, $\alpha^{2}F(\omega)$,
were calculated using the density-functional perturbation theory \cite{baroni_rmp}
as implemented in the PWSCF package. Finally, the isotropic Eliashberg
gap equation \cite{allen1,allen2,private1} has been solved for a
range of $\mu^{*}$ to obtain the corresponding $T_{c}.$ 

The charge self-consistent, full-potential, LMTO calculations for
$MgB_{2}$ were carried out using the local-density approximation
for exchange-correlation of Perdew \emph{et al}. \cite{perdew1},
$3\kappa$-energy panels, and 793 $\mathbf{k}$-points in the irreducible
wedge of the Brillouin zone (BZ).  The 2$p$ state of $Mg$ was treated
as a semi-core state. The basis set used consisted of $s$, $p$,
and $d$ orbitals at the $Mg$ and the $B$ sites, and the potential
and the wave function were expanded up to $l_{max}=6$. The muffin-tin
spheres for $Mg$ and $B$ were taken to be slightly smaller than
the touching spheres in each case. These calculations are used to
study the symmetry-resolved densities of states as well as the band-structure
along the high symmetry directions in the corresponding BZ. 

For calculating the Fermi surface, the phonon dispersion, the phonon
density of states, and the Eliashberg function, we have used the plane-wave
pseudopotential method with Vanderbilt's ultrasoft pseudopotentials
with nonlinear core correction. The kinetic energy cutoff for the
wave function was 35 Ry, and for the charge density and the potential
it was 200 Ry. The exchange-correlation potential was parametrized
as suggested by Perdew \emph{et al}. \cite{perdew1}. The Fermi surface
was constructed using eigenvalues calculated on a $36\times36\times36$
grid in the reciprocal space. The linear-response calculations were
carried out on a $6\times6\times6$ grid resulting in $28$ irreducible
$\mathbf{q}$-points. A $12\times12\times12$ grid of $\mathbf{k}$-points
is used for the BZ integrations during linear response calculations.
The electron-phonon matrix elements are sampled more accurately on
a $36\times36\times36$ grid using the double-grid technique. To check
the convergence of our linear response calculations, we have used
an $18\times18\times18$ and $24\times24\times24$ grids for carrying
out the BZ integrations during self-consistency for points along $\Gamma-A$
direction as well as for pressures above $125$ GPa. The convergence
of electron-phonon matrix elements were checked with grids up to $40\times40\times40$
in the reciprocal space. 

\begin{figure}
\begin{centering}\includegraphics[clip,scale=0.33]{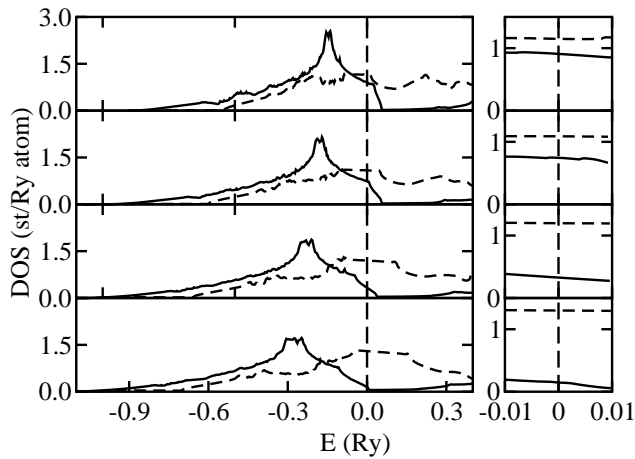}\par\end{centering}

\caption{The $B$ $p_{x(y)}$(solid line) and the $p_{z}$ (dashed line) densities
of states in $MgB_{2}$ as a function of pressure for $0$ ($1$st
panel from the top), $50$ ($2$nd panel), $100$ ($3$rd panel),
and $137$ ($4$th panel ) GPa. The panels on the right show the corresponding
densities of states in a $10$ mRy energy window around the Fermi
energy, which is shown as the vertical, long-dashed line in the figure.
\label{dos}}
\end{figure}

The $B$ $p_{x(y)}$ and $p_{z}$ densities of states (DOS) play a
crucial role in determining the superconducting properties of $MgB_{2}$.
Our calculated $B$ $p_{x(y)}$ and $p_{z}$ DOS as a function of
pressure are shown in Fig. \ref{dos}, where we also show the DOS
in a $10$ mRy energy window around Fermi energy, $E_{F}$, in a separate
panel. With increasing pressure both $p_{x(y)}$ and $p_{z}$ DOS
flatten and the states are pushed further down. Since with increasing
pressure the lattice constant $c$ decreases much more rapidly than
$a$ in $MgB_{2}$, as a result the $p_{x(y)}$ DOS around $E_{F}$
is significantly reduced. Between $100$ and $137$ GPa the $p_{x(y)}$
DOS is inside $E_{F}$ with essentially no holes. However, the $p_{z}$
DOS around $E_{F}$ is in tact up to $137$ GPa. Such a situation
is very different from the usual band filling that takes place when
we put $Al$ and/or $C$ as impurities in $MgB_{2}$. These differences
can be further highlighted by considering the changes in the band-structure
along high symmetry directions in the BZ of $MgB_{2}$ as a function
of pressure.

\begin{figure}
\begin{centering}\includegraphics[clip,scale=0.33,angle=270]{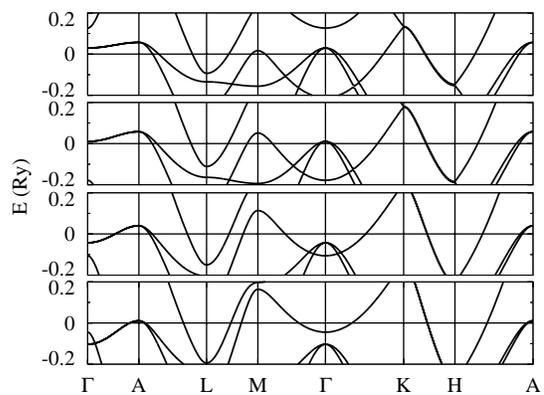}\par\end{centering}

\begin{centering}\par\end{centering}

\caption{The band-structure of $MgB_{2}$ along high symmetry directions in
the Brillouin zone as a function of pressure for $0$ ($1$st panel
from the top), $50$ ($2$nd panel), $100$ ($3$rd panel), and $137$
($4$th panel) GPa. The horizontal line, passing through the energy
zero, indicates the Fermi energy. Note that the numerical values of
the wave vector $\mathbf{k}$ corresponding to the various symmetry
points, as shown in the figure, are different at different pressures.
\label{band}}
\end{figure}

In Fig. \ref{band}, we show the calculated band-structure of $MgB_{2}$
along high symmetry directions in the Brillouin zone. The states along
$\Gamma$-A have $B$ $p_{x(y)}$ character while states at $M$ are
of $B$ $p_{z}$ type. With increasing pressure, as can be seen from
Fig. \ref{band}, the states along $\Gamma$-$A$ move towards $E_{F}$
while the states at $M$ move away from $E_{F}$. By $137$ GPa, all
the $B$ $p_{x(y)}$-derived states along $\Gamma$-$A$ are close
to being inside $E_{F}$, while the states at $M$ have moved up by
almost $0.17$ Ry. The movement of states along $\Gamma$-$A$ and
$M$ with increase in pressure, in conjunction with doping of $MgB_{2}$,
can be more effective in tuning its superconducting properties. 

\begin{figure}
\begin{centering}\subfigure[]{\includegraphics[clip,scale=0.25]{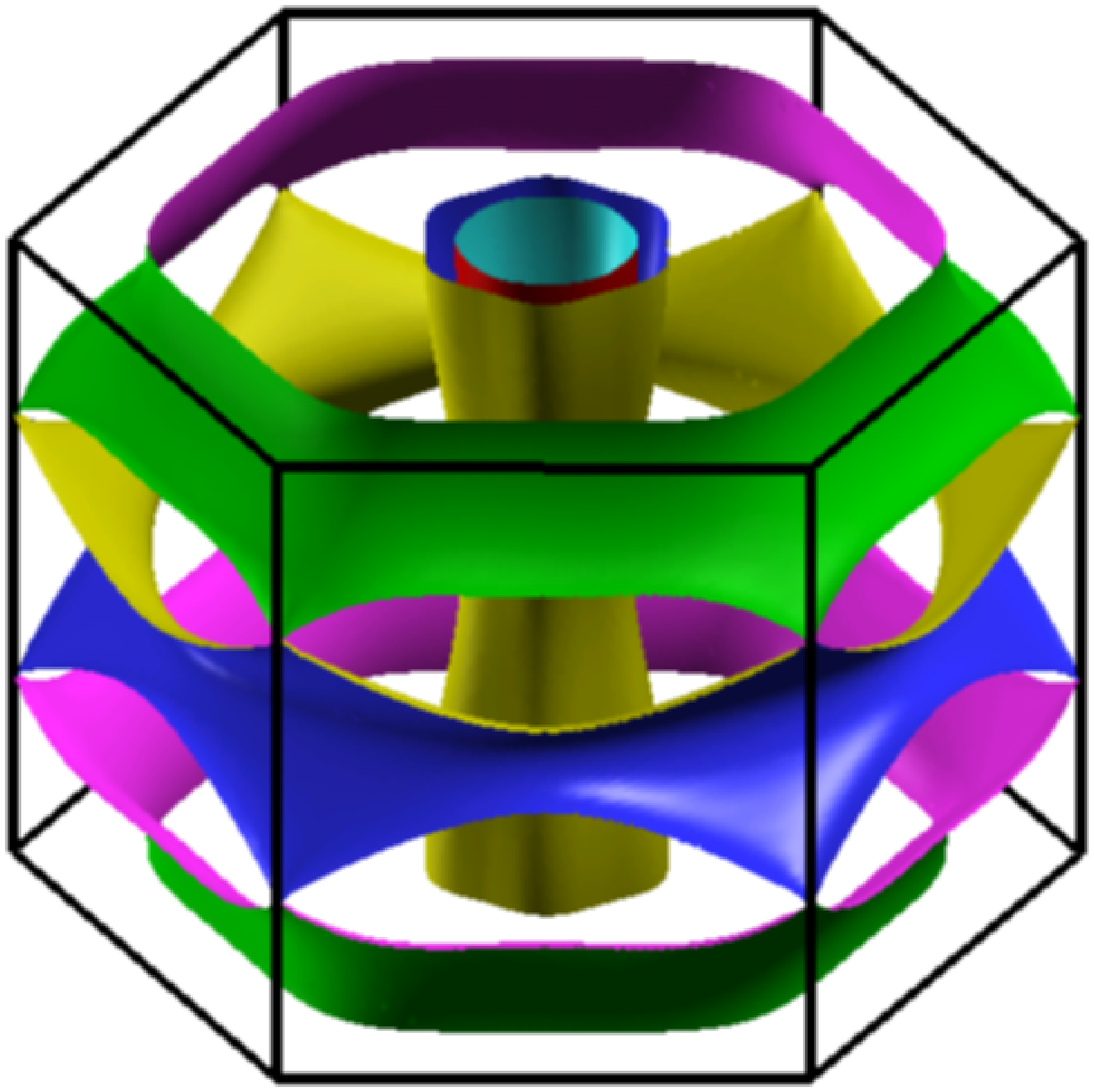}}~
\subfigure[]{\includegraphics[scale=0.25]{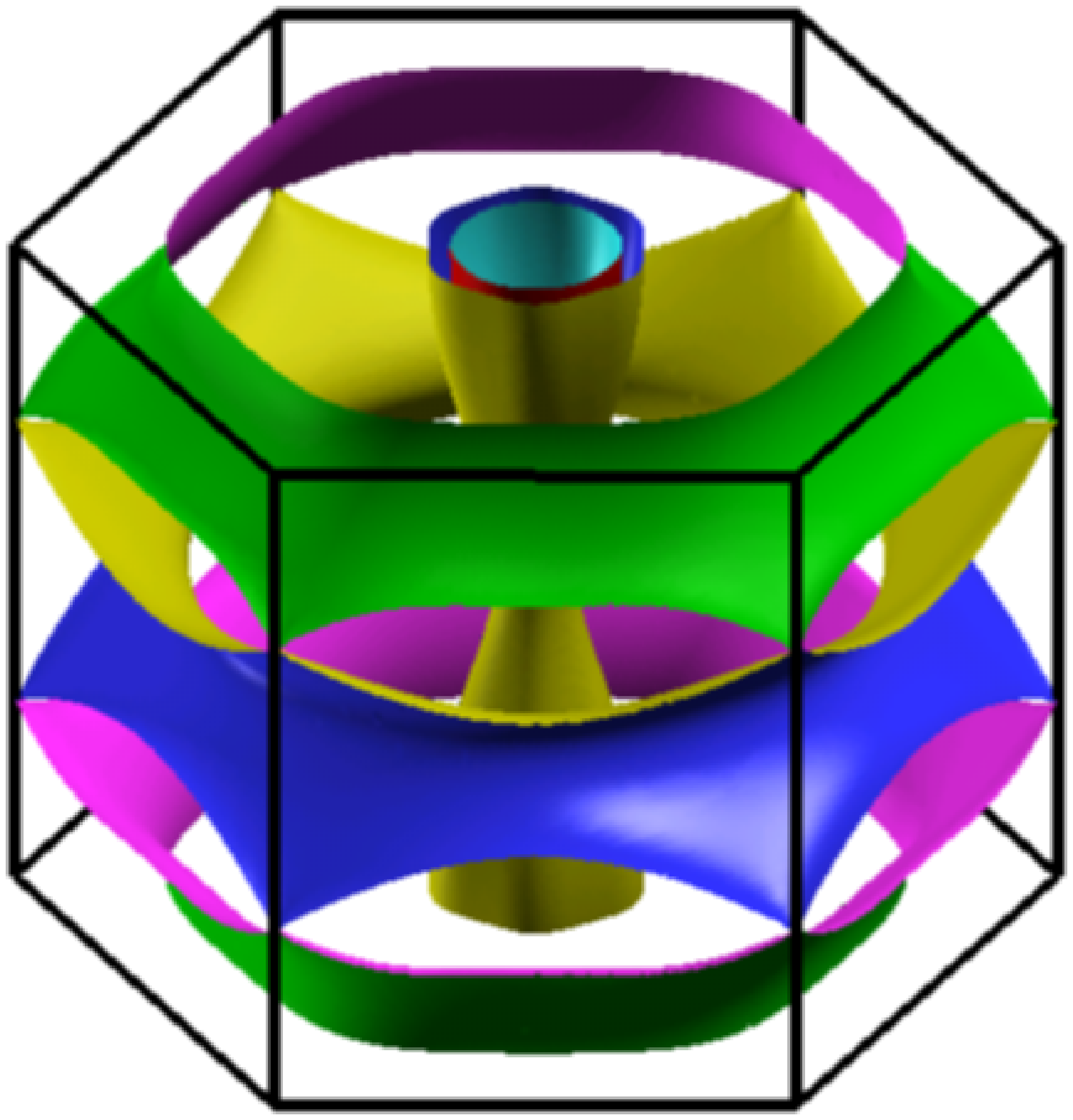}} \par\end{centering}

\begin{centering}\subfigure[]{\includegraphics[scale=0.25]{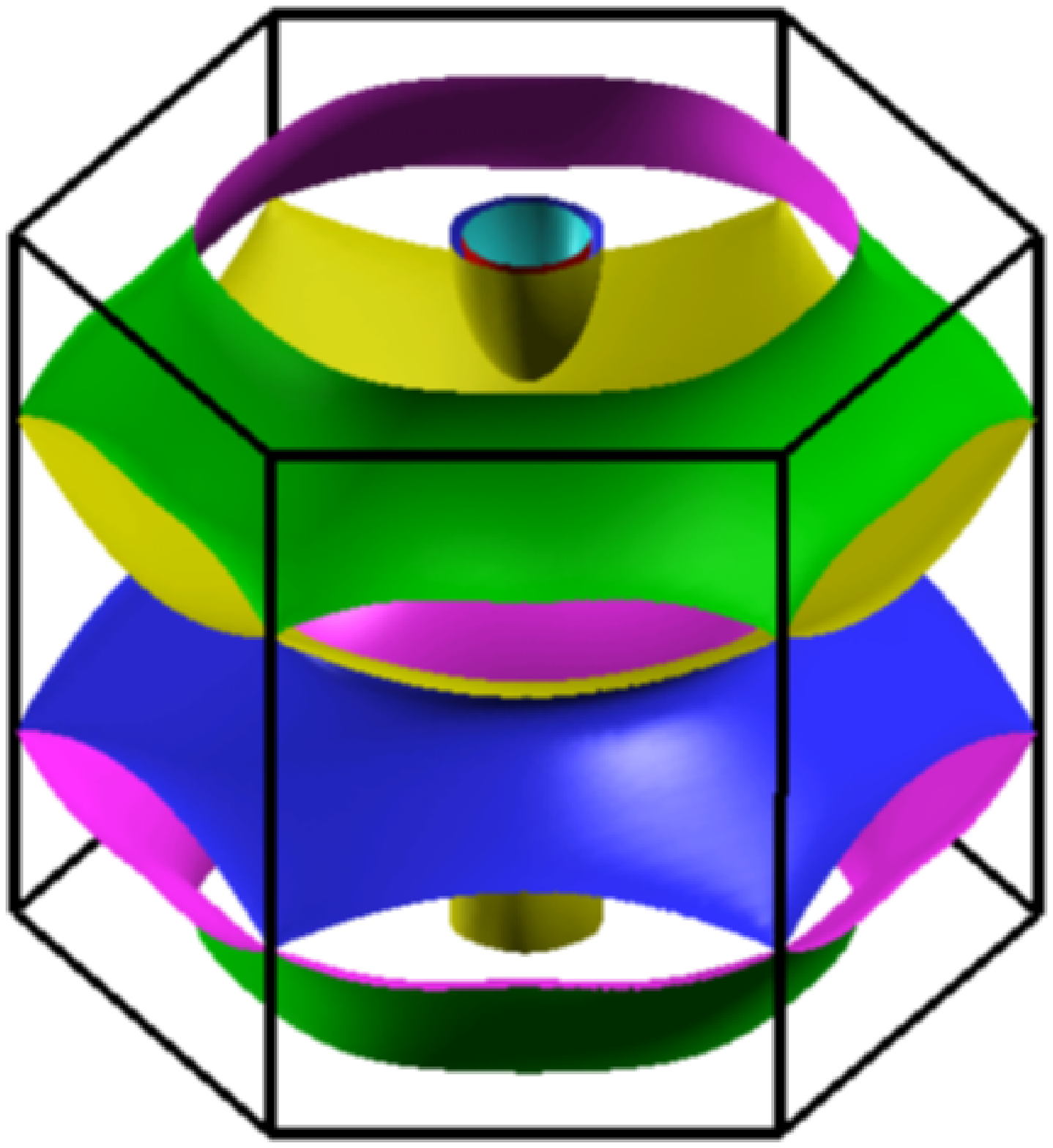}}
~\subfigure[]{\includegraphics[scale=0.23]{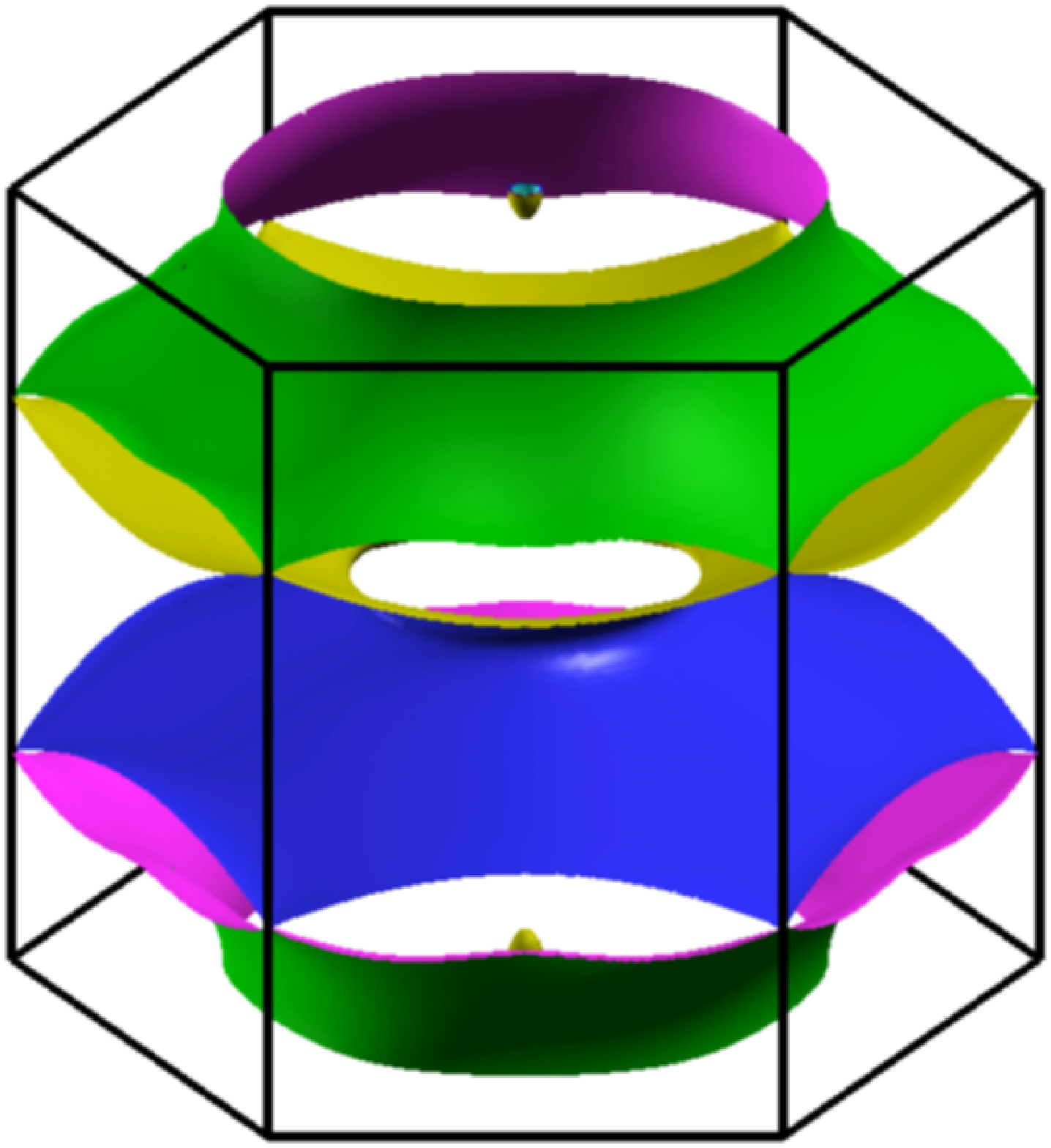}}\par\end{centering}

\caption{The Fermi surface of $MgB_{2}$ as a function of pressure in the
corresponding Brillouin zone for (a) $0$, (b) $50$, (c) $100$ and
(d) $137$ GPa, calculated as described in the text. \label{fs}}
\end{figure}

A complete picture of the pressure-induced changes in the electronic
structure of $MgB_{2}$ emerges through the changes in the Fermi surface,
as shown in Fig. \ref{fs}. We find that the characteristic two-dimensional
cylindrical sheets along $\Gamma$-$A$ become narrower and constricted
around $\Gamma$ point with increase in pressure. With further increase
in pressure, say by $100$ GPa, the states manage to form only a small
portion of the tube around $A$ point of the $\Gamma$-$A$ direction.
The circular opening of the $p_{z}$-derived Fermi surface sheets
containing ($K,$$M$) and ($H,$$L$) symmetry points narrows as
the band moves up. At a pressure of $137$ GPa, the Fermi surface
consists of essentially these sheets with the cylindrical tube reduced
to a dot around $A$ point. Thus, as long as the states on the cylindrical
sheets play a pivotal role in sustaining superconductivity in $MgB_{2}$,
their absence beyond $100$ GPa, as shown in Fig. \ref{fs}, ensures
the vanishing of superconductivity in $MgB_{2}$ around $100$ GPa
and above. Unless, of course, with pressures above $100$ GPa the
electron-phonon coupling provided by the remaining states is strong
enough to \emph{restart} the superconductivity in $MgB_{2}$. 

\begin{figure}
\begin{centering}\includegraphics[clip,scale=0.33,angle=270]{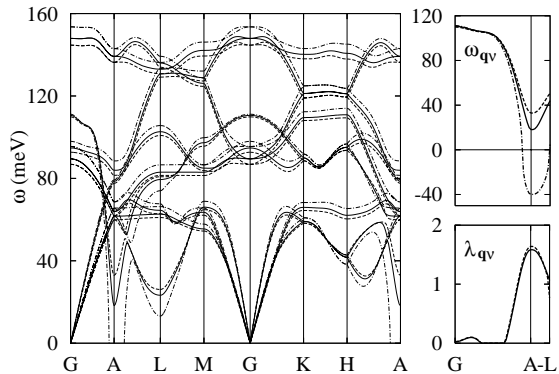}\par\end{centering}

\begin{centering}\par\end{centering}

\caption{The phonon dispersion $\omega$ of $MgB_{2}$ along high symmetry
directions in the Brillouin zone (left panel), and the $\omega_{\mathbf{q}\nu}$
(top-right panel) and the $\gamma_{\mathbf{q}\nu}$ (bottom-right
panel) for $B_{1g}$ phonon mode along $\Gamma$-$A$-$L$, as a function
of pressure for $125$ (dashed line), $137$ (solid line) and $143$
(dot-dashed line) GPa. The $\omega_{\mathbf{q}\nu}$ and the $\gamma_{\mathbf{q}\nu}$
are in meV, the frequencies below zero (in the top-right panel) are
imaginary, and for $143$ GPa the $\gamma_{\mathbf{q}\nu}$ is not
shown. Note that the numerical values of $\mathbf{q}$ corresponding
to the various symmetry points are different at different pressures.
\label{phonon}}
\end{figure}

To be able to go beyond the qualitative assessments that we have made
so far about the pressure-induced changes in the electron-phonon interaction
in $MgB_{2}$ using the changes in the electronic structure as the
backdrop, we have calculated the phonon dispersion, the phonon linewidth,
the phonon density of states and the electron-phonon interaction as
a function of pressure up to $150$ GPa. We find that in $MgB_{2}$
at zero pressure the frequency associated with $E_{2g}$ phonon mode
at $\Gamma$ is $67$ meV, which goes up to $148$ meV at $137$ GPa.
Similarly, the frequency of the $E_{2u}$ phonon mode at $A$ point
increases from $60$ meV at zero pressure to $139$ meV at $137$
GPa. In contrast, the frequency of the $B_{1g}$ phonon mode at $A$
point, involving out-of-plane motion of $B$ atoms, reduces from $63$
meV at $100$ GPa to $34$ meV at $125$ GPa. As shown in Fig. \ref{phonon},
the softening of $B_{1g}$ mode around $A$ point continues with further
increase in pressure and by $143$ GPa it has become unstable. The
softening of $B_{1g}$ mode around $A$ point is accompanied by an
enhanced $\gamma_{\mathbf{q}\nu}$, as can be seen from Fig. \ref{phonon}.
Note, however, that the dynamical instability of $B_{1g}$ mode in
the present calculation does not necessarily imply a physical instability
of the lattice. Under these conditions, the anharmonic effects become
important. 

\begin{figure}
\begin{centering}\includegraphics[clip,scale=0.33]{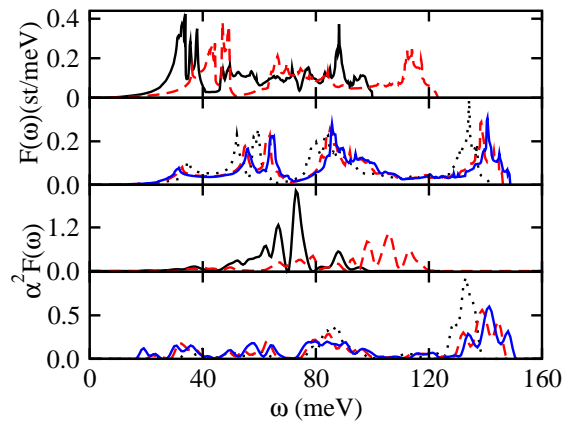}\par\end{centering}

\caption{The phonon density of states $F(\omega)$ (top 2 panels) and the
Eliashberg function $\alpha^{2}F(\omega)$ (bottom 2 panels) of $MgB_{2}$
as a function of pressure. The $F(\omega)$ and $\alpha^{2}F(\omega)$
for $0$ (solid line) and $50$ (dashed line) GPa are shown in the
$1$st panel and the 3rd panel, respectively. The $F(\omega)$ and
$\alpha^{2}F(\omega)$ for $100$ (dotted line), $125$ (dashed line)
and $137$ (solid line) GPa are shown in the $2$nd panel and the
4th panel, respectively. \label{ep}}
\end{figure}

\begin{figure}
\begin{centering}\includegraphics[clip,scale=0.33]{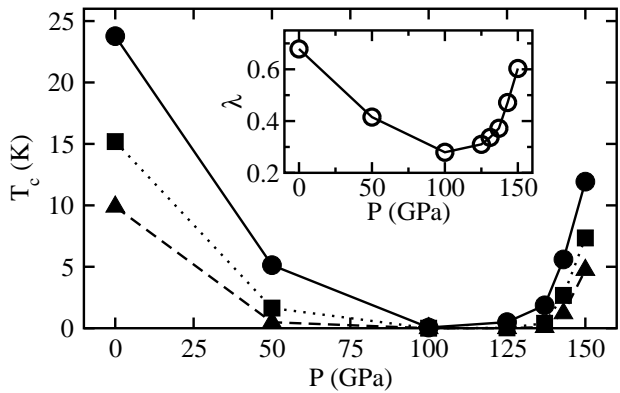}\par\end{centering}

\caption{The calculated superconducting transition temperature $T_{c}$ in
$MgB_{2}$ as a function of pressure $P$ for $\mu^{*}$ equal to
$0.1$ (solid circles connected with solid line), $0.15$ (solid squares
connected with dotted line) and $0.2$ (solid triangles connected
with dashed line), respectively. The inset shows the variation of
$\lambda$ with pressure. The $\lambda$ and  $T_{c}$ at $143$ and
$150$ GPa are obtained by ignoring the coupling of imaginary frequencies.
\label{tc}}
\end{figure}

Our calculated $F(\omega)$ and $\alpha^{2}F(\omega)$ of $MgB_{2}$
for pressures up to $137$ GPa are shown in Fig. \ref{ep}. For $MgB_{2}$
at zero pressure, our results for $F(\omega)$ compare well with previous
calculations \cite{bohnen,kong,choi1,yildirim}. The decrease in the
strength of the electron-phonon coupling in $MgB_{2}$ with increase
in pressure can be clearly seen in Fig. \ref{ep}. The average electron-phonon
coupling constant $\lambda$, shown in Fig. \ref{tc}, decreases from
$0.68$ to $0.28$ as the pressure is increased from zero to $100$
GPa, which is consistent with the expected loss of superconductivity
in $MgB_{2}$ by this pressure. By $100$ GPa the partial $\lambda_{q\nu}$
of the $E_{2g}$ phonon mode at $\Gamma$ and $A$ points are reduced
by a factor of $3$ and $20$, respectively. However, the increase
in $\lambda$ to $0.38$ by $137$ GPa with the possibility of higher
values with further increase in pressure, as shown in Fig. \ref{tc},
suggests the reemergence of superconductivity in $MgB_{2}$ at higher
pressures \emph{without} the dominance of $E_{2g}$ mode. 

Finally, the change in $T_{c}$ with pressure has been obtained by
solving the isotropic gap equation \cite{allen2,private1} using the
calculated Eliashberg function $\alpha^{2}F(\omega)$ at each pressure.
The results of these calculations are shown in Fig. \ref{tc} for
a range of $\mu^{*}$. Our results, shown in Fig. \ref{tc}, predict
the reemergence of superconductivity in $MgB_{2}$ at pressures above
$100$ GPa. In particular, we find that at $137$ GPa $MgB_{2}$ will
superconduct with $T_{c}\approx2$ K for $\mu^{*}=0.1$. Since the
electron-phonon interaction in $MgB_{2}$ at zero pressure is known
to be very anisotropic, an accurate determination of $T_{c}$ requires
the solution of the anisotropic gap equation. However, we have seen
that the $B$ $p_{x(y)}$-dependent electron-phonon coupling decreases
dramatically with increasing pressure, making the use of isotropic
gap equation for the calculation of $T_{c}$ more reliable at higher
pressures. 

In conclusion, we have studied the effects of pressure on the electron-phonon
interaction in $MgB_{2}$ using density-functional-based methods.
The increase in pressure hardens the lattice and dramatically reduces
the contribution of the $E_{2g}$ phonon mode to the electron-phonon
coupling. As a result, the superconductivity in $MgB_{2}$ vanishes
by $100$ GPa, only to \emph{reappear} at higher pressures. In particular,
we find a superconducting transition temperature $T_{c}\approx2$
K for $\mu^{*}=0.1$ at a pressure of $137$ GPa. 

\end{document}